\documentclass[aps,PRB,twocolumn,superscriptaddress,preprintnumbers]{revtex4-2}
\usepackage[T1]{fontenc}
\usepackage{times}
\usepackage{graphicx,color}
\usepackage{amsfonts,amsmath,amssymb,amsbsy}
\usepackage[colorlinks=true, linkcolor=blue, citecolor=blue, urlcolor=blue]{hyperref}
\usepackage{float}

\begin{document}

\title{ Topological superconductors and Majorana fermions \\
based on $X$-wave magnets with $X=p,d,f,g,i $}
\author{Motohiko Ezawa}
\affiliation{Department of Applied Physics, The University of Tokyo, 7-3-1 Hongo, Tokyo
113-8656, Japan}

\begin{abstract}
We study superconductors coupled with $X$-wave magnets with $X=p,d,f,g,i$.
They are essentially different between altermagnets with $X=d,g,i$ and
odd-parity magnets $X=p,f$ due to the compatibility of the spin-singlet
formation of the $s$-wave superconductor. The altermagnets and odd-parity
magnets show rich superconducting phase diagrams depending on the strength
of the magnetic order, the chemical potential and the ratio of the on-site
and nearest-neighbor interactions. Especially, we find that the
altermagnetic order stabilizes the chiral ($p_{x}+ip_{y}$)-wave
superconductor, which is a class D topological superconductor hosting
Majorana chiral edge states. It is characterized by the Chern number.
Especially, there emerge two chiral edge modes in the $i$-wave altermagnet,
which are characterized by the Chern number 2. On the other hand, the $p$ ($f$%
)-wave odd-parity magnetic order stabilizes the $p$ ($f$)-wave topological
superconductor, which is a class DIII topological superconductor hosting
Majorana flat bands. It is characterized by the winding number. Majorana
flat bands are robust in the presence of the mixing with the $s$-wave
superconductivity, which is inevitable for odd-parity magnets although the
magnitude of the mixing is tiny.
\end{abstract}

\date{\today }
\maketitle

\section{Introduction}

Superconductivity is one of the most fascinating phenomena in
condensed-matter physics. A conventional superconductor has the $s$-wave
superconducting gap, which is isotropic and has a full gap. On the other
hand, $d$-wave superconductivity is well studied in the context of the
high-temperature superconductivity. In addition, $p_{x}+ip_{y}$
superconductors are known to be topological superconductors\cite{Volovik},
which have Majorana chiral-edge states.

Altermagnets\cite{SmejX,SmejX2,Naka,Gonza,NakaB,Bose,NakaRev,GI} and
odd-parity magnets\cite%
{Hayami,pwave,He,Comin,Yamada,HZhou,Brek,EzawaPNeel,Chak,Atasi,PEdel} are
new-types of magnets with zero-net magnetization, which are characterized by
the spin-split electronic band structure. They are protected by a
combinatorial symmetry of the spin flip and the momentum rotation.
Altermagnets constitute of $d$-wave, $g$-wave and $i$-wave magnets, while
odd-parity magnets constitute of $p$-wave and $f$-wave magnets. They are
summarized as $X$-wave magnets\cite{Planar,MTJ,APEX} with $X=p,d,f,g,i$. The
interplay between superconductivity and $X$-wave magnets is interesting
because there is no net magnetization in the $X$-wave magnet. It is shown
that the chiral $p_{x}+ip_{y}$-wave topological superconductor is stabilized when it is 
coupled with the $d$-wave altermagnet\cite{Hong}. On the other hand, the
Majorana flat bands emerge when it is coupled with the $p$-wave magnet\cite{KYKim},
where there is a mixing of the $p_{x}$-wave and $s$-wave superconducting
gaps. Although superconductors coupled with $d$-wave altermagnets\cite%
{Zhu,Li23,Ghora,Chak25,Chat,Maeda,FukayaRev,Alam,HHu,Xiao,Jas} and $p$-wave
magnets\cite{EzawaPwave,Suk,Kho,Pal,KYKim,Luo,Carmero} are studied, those
coupled with $g$-wave altermagnets, $i$-wave altermagnets and $f$-wave
altermagnets have been scarcely studied.

In this paper, we study superconductors coupled with $X$-wave magnets. We
find that the phase diagrams are essentially different between altermagnets
and odd-parity magnets. The difference arises from the fact that
spin-singlet superconducting pairing is compatible with the altermagnetic
order, while it is not in odd-parity magnets. Especially, the $p$-wave
chiral topological superconductor is realized in the case of altermagnets.
Furthermore, chiral ($p_{x}+ip_{y}$) superconductivities emerge universally
in the case of the $d$-wave, $g$-wave and $i$-wave altermagnets. They are
class D topological superconductors\cite{Schnyder} due to the time-reversal
breaking of altermagnets. There emerge chiral edge modes, which are
characterized by the Chern number. Especially, there emerge two chiral edge
modes in the $i$-wave altermagnet, which are characterized by the Chern
number 2. On the other hand, there emerge Majorana flat bands in the $p$%
-wave ($f$-wave) superconductor coupled with the $p$-wave ($f$-wave) magnet,
which is characterized by the modulo 2 of the winding number. They are class
DIII topological superconductors\cite{Schnyder} because odd-parity magnets
preserve time-reversal symmetry in the electronic band structure.

\section{Results}

\subsection{Superconducting gap}

We study the extended Hubbard model with on-site and nearest-neighbor
attractive interactions,%
\begin{align}
& H_{\text{Hubbard}}  \notag \\
=& -t\sum_{\left\langle j,j^{\prime }\right\rangle ,s}c_{js}^{\dagger
}c_{j^{\prime }s}-U\sum_{j}c_{j\uparrow }^{\dagger }c_{j\uparrow
}c_{j\downarrow }^{\dagger }c_{j\downarrow }-V\sum_{\left\langle j,j^{\prime
}\right\rangle }n_{j}n_{j^{\prime }}  \label{Hubbard}
\end{align}%
with the density%
\begin{equation}
n_{j}=\sum_{s}c_{js}^{\dagger }c_{js},
\end{equation}%
where $t$ denotes the hopping amplitude, $U$ is the strength of the on-site
attractive interaction strength, and $V$ represents the attractive
interaction strength between nearest-neighbor sites. $\left\langle
j,j^{\prime }\right\rangle $ represents the nearest-neighbor sites $j$ and $%
j^{\prime }$, and $c_{js}^{\dagger }$ ($c_{js}$) represents the creation
(annihilation) operator of electrons with spin $s=\uparrow ,\downarrow $. We
make an analysis based on the square lattice for $d$-wave altermagnets, $g$%
-wave altermagnets and $p$-wave magnets, while the triangular lattice for $i$%
-wave altermagnets and $f$-wave magnets.

Due to the Pauli exclusion principle, spin-singlet superconducting gaps must
satisfy 
\begin{equation}
\Delta \left( -\mathbf{k}\right) =\Delta \left( \mathbf{k}\right) ,
\end{equation}%
while spin-triplet superconducting gaps must satisfy%
\begin{equation}
\Delta \left( -\mathbf{k}\right) =-\Delta \left( \mathbf{k}\right) .
\end{equation}%
Spin-singlet superconducting gaps have $s$-wave, $d$-wave, $g$-wave and $i$%
-wave symmetries, which have even parities. On the other hand, spin-triplet
superconducting gaps have $p$-wave and $f$-wave symmetries, which have odd
parities. We note that $g$-wave and $i$-wave superconducting gaps cannot be
stabilized in the Hamiltonian (\ref{Hubbard}) because they require
longer-range interactions. Furthermore, $f$-wave superconducting gap is
realized only in the triangular lattice.

We study the superconductivity based on the Bogoliubov-de Gennes (BdG) Hamiltonian,%
\begin{equation}
H_{\text{BdG}}\left( \mathbf{k}\right) =%
\begin{pmatrix}
\hat{\xi}(\mathbf{k}) & \Delta (\mathbf{k}) \\ 
\Delta ^{\dagger }(\mathbf{k}) & -\hat{\xi}^{T}(-\mathbf{k})%
\end{pmatrix}%
,
\end{equation}%
where $\hat{\xi}(\mathbf{k})$ is the kinetic energy. We solve the linearized
gap equation and calculate which superconductor has the highest critical
temperature, and determine the phase diagram. Details are shown in Methods %
\ref{GapEq}.

\begin{figure}[t]
\centerline{\includegraphics[width=0.48\textwidth]{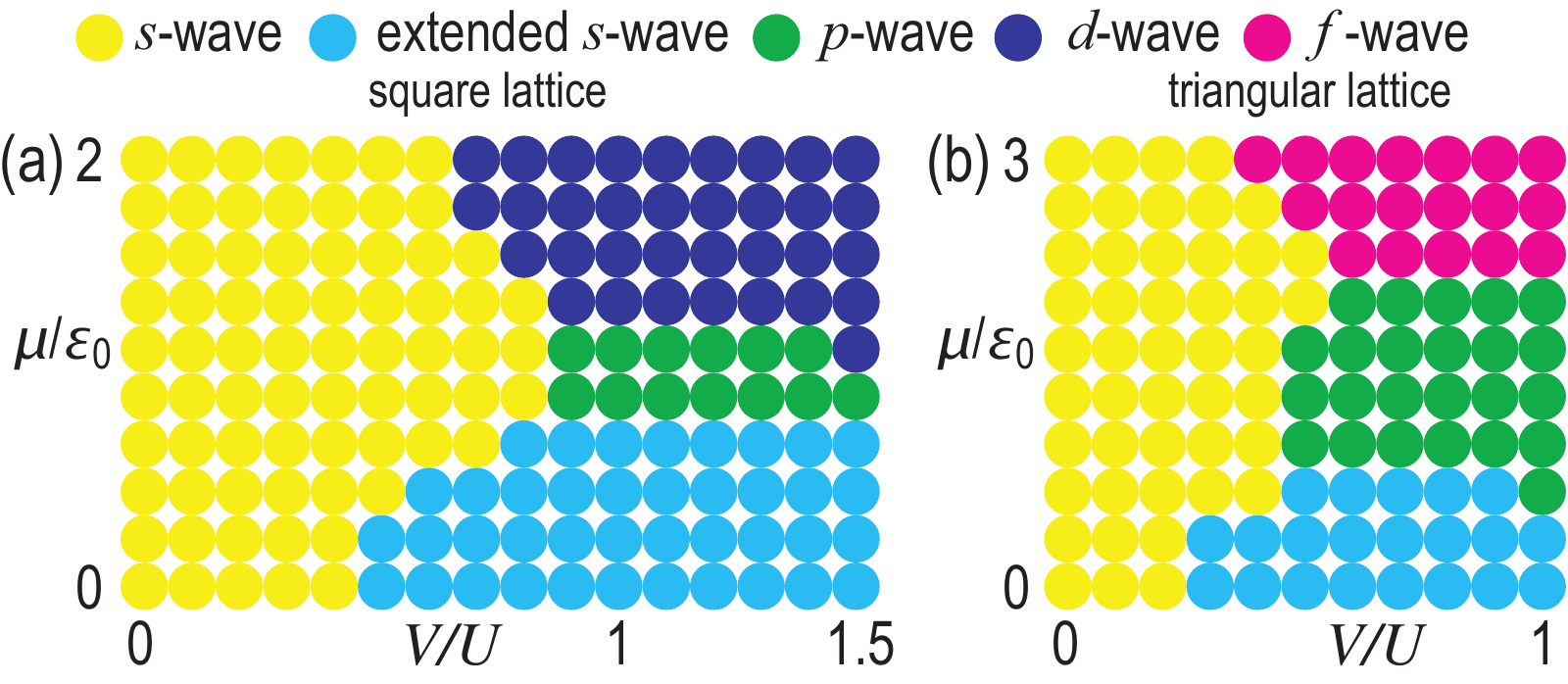}}
\caption{Phase diagram with $J=0$. (a) Square lattice and (b) triangular
lattice. The horizontal axis is $V/U$ and the vertical axis is $\protect\mu /%
\protect\varepsilon _{0}$. Yellow disks represent the $s$-wave
superconductivity, cyan disks represent the extended $s$-wave
superconductivity, green disks represent the in-plane chiral $p$-wave
superconductivity in (a) and out-of-plane $p_{x}$-wave superconductivity in
(b), blue disks represent the $d$-wave superconductivity, and magenta disks
represent the $f$-wave superconductivity. We have set $t=\protect\varepsilon %
_{0}/2$ and $U=\protect\varepsilon _{0}$, where $\protect\varepsilon _{0}$
is the unit of energy.}
\label{FigMu}
\end{figure}

\subsection{Phase diagram without magnetic order}

We start with a review of the phase diagram of superconductors without the $%
X $-wave magnet. When the on-site interaction is dominant, the $s$-wave
superconductor is realized. On the other hand, the phase diagram shows a
rich variety when the nearest-neighbor interaction is dominant. It depends
on whether the lattice is square or triangular.

We first study the square lattice, whose phase diagram is shown in Fig.\ref%
{FigMu}(a). The extended $s$-wave superconductor induced by $V$ is realized
in the vicinity of the band bottom, where the Fermi surface is almost a
circle. See Eq.(\ref{exs}) for the definition of the gap function of the
extended $s$-wave superconductivity. The chiral $p$-wave superconductor is
realized when the chemical potential is near the 1/4 filling. The $d$-wave
superconductor is realized at the half filling.

We next study the triangular lattice, whose phase diagram is shown in Fig.%
\ref{FigMu}(b). The extended $s$-wave superconductor is realized for $%
0.3\lesssim V/U\lesssim 0.6$, while the chiral $p$-wave superconductor is
realized for $V/U\gtrsim 0.6$ in the vicinity of the band bottom. The chiral 
$p$-wave superconductor is realized when the chemical potential is near the
1/4 filling. The $f$-wave superconductor is realized when $V$ is dominant.
The phase boundary $V/U$\ between the $s$-wave superconductor and the
extended $s$-wave superconductor in the triangular lattice is smaller than
that in the square lattice. It is because the number of the adjacent site is
six in the triangular lattice and it is four in the square lattice, where
the effective next-nearest neighbor interactions in the triangular lattice
are stronger than those in the square lattice.

\begin{figure*}[t]
\centerline{\includegraphics[width=0.88\textwidth]{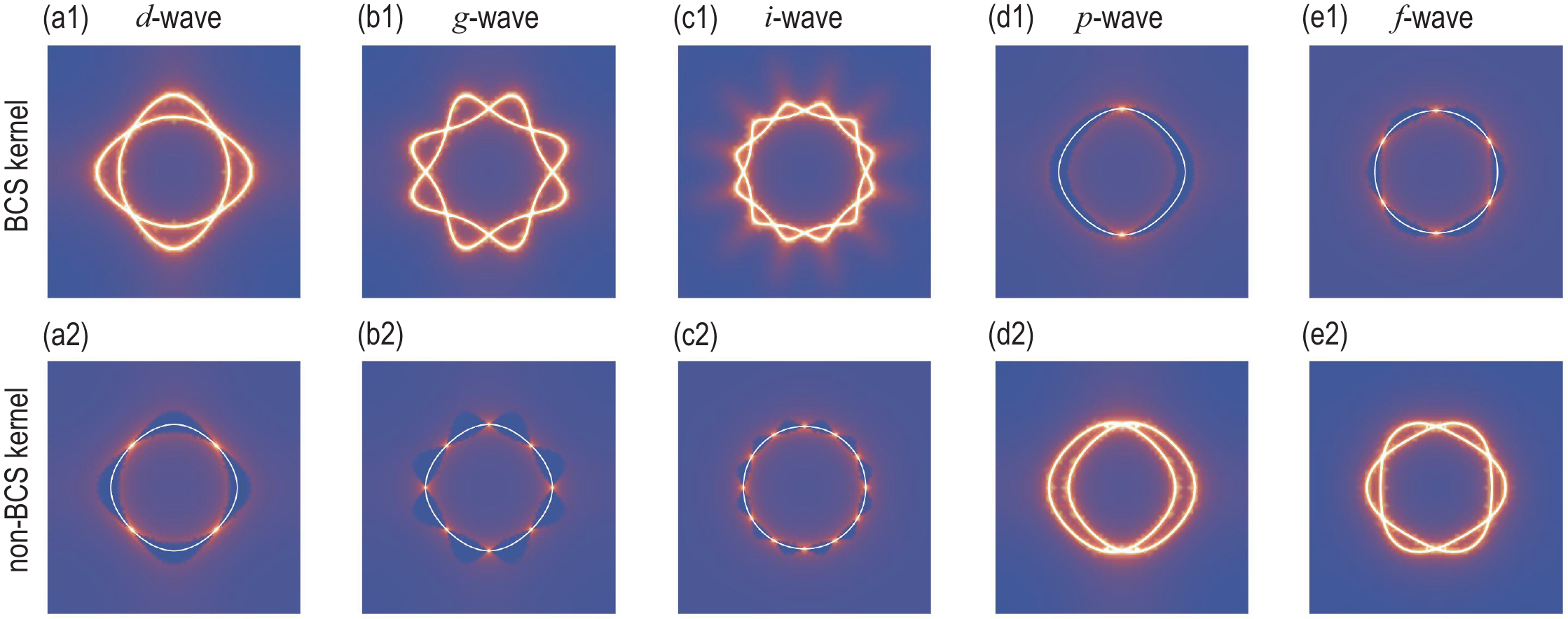}}
\caption{{}(a1)$\sim $(e1) Momentum distribution of the BCS kernel $K_{\text{%
BCS}}\left( \mathbf{k}\right) $. (a2)$\sim $(e2) Momentum distribution of
the non-BCS kernel $K_{\text{BCS}}^{\text{no}}\left( \mathbf{k}\right) $.
The horizontal axis and the vertical axis is $k_{x}$ and $k_{y}$,
respectively. We have set $\protect\mu /\protect\varepsilon _{0}=1$. }
\label{FigDensity}
\end{figure*}

\subsection{$X$-wave magnets}

The electromagnetic property of the $X$-wave magnet is characterized by the
peculiar band splitting depending on the spin. The simplest model is the
two-band Hamiltonian\cite{GI,Planar,MTJ,APEX,NLSeebeck} given by\cite%
{Xiao,KYKim,Hong,Carmero,HHu}%
\begin{equation}
H_{X}=H_{\text{Hubbard}}\left( \mathbf{k}\right) +Jf_{X}\left( \mathbf{k}%
\right) \sigma _{z},
\end{equation}%
where the first term $H_{\text{Hubbard}}\left( \mathbf{k}\right) $
represents the Hamiltonian (\ref{Hubbard}), while the second term represents
the\ band splitting described by the function $f_{X}\left( \mathbf{k}\right) 
$\ with the coupling constant $J$\ and the Pauli matrix $\sigma _{z}$.
Explicit forms of $f_{X}\left( \mathbf{k}\right) $ are given in Methods \ref%
{tight}. We have $f_{X}\left( -\mathbf{k}\right) =f_{X}\left( \mathbf{k}%
\right) $ for altermagnets, while $f_{X}\left( -\mathbf{k}\right)
=-f_{X}\left( \mathbf{k}\right) $ for odd-parity magnets. As a result, the
linearized gap equation for altermagnets reads%
\begin{equation}
-\frac{V_{\alpha \alpha }\,k_{\text{B}}T}{N_{\Gamma }}\sum_{\mathbf{k}}\frac{%
\Gamma _{\alpha }^{\dagger }\Gamma _{\alpha }}{4H_{\text{kine}}\left( 
\mathbf{k}\right) k_{\text{B}}T}\sum_{s=\pm 1}\tanh \frac{\varepsilon
_{s}\left( \mathbf{k}\right) }{2k_{\text{B}}T}=1  \label{GapEqA}
\end{equation}%
for the singlet superconducting gap and the out-of-plane triplet
superconducting gap, while it reads%
\begin{equation}
-\frac{V_{\alpha \alpha }\,k_{\text{B}}T}{N_{\Gamma }}\sum_{\mathbf{k}}\frac{%
\Gamma _{\alpha }^{\dagger }\Gamma _{\alpha }}{4k_{\text{B}}T}\sum_{s=\pm 1}%
\frac{\tanh \frac{\varepsilon _{s}\left( \mathbf{k}\right) }{2k_{\text{B}}T}%
}{\varepsilon _{s}\left( \mathbf{k}\right) }=1  \label{GapEqB}
\end{equation}%
for the in-plane triplet superconducting gap. For odd-parity magnets, Eq.(%
\ref{GapEqB}) is the gap equation for the singlet superconducting gap and
the out-of-plane triplet superconducting gap, while Eq.(\ref{GapEqA}) is the
gap equation for the in-plane triplet superconducting gap. Namely, the gap
equations are reversed for the altermagnets and the odd-parity magnets. See
details in Methods \ref{GapEq}.

\subsection{Altermagnets}

Altermagnets break time-reversal symmetry. It leads to the class D
topological superconductors with the $p_{x}+ip_{y}$ chiral superconducting
gap, which is a full gap. We show that there emerge chiral edge states,
which are characterized by the Chern number.

The critical temperature of the solution of Eq.(\ref{GapEqA}) parabolically
decreases as the increase of $J$ as%
\begin{equation}
T_{\text{cr}}=T_{\text{cr}}^{\left( 0\right) }-28\zeta ^{\prime }\left(
-2\right) \frac{J^{2}\left\langle f\left( \mathbf{k}\right)
^{2}\right\rangle _{\text{FS}}}{T_{\text{cr}}^{\left( 0\right) }},
\label{Tor}
\end{equation}%
where $\left\langle f\left( \mathbf{k}\right) ^{2}\right\rangle _{\text{FS}}$
is the integration over the Fermi surface, and%
\begin{equation}
\zeta ^{\prime }\left( x\right) \equiv \frac{d\zeta \left( x\right) }{dx}
\end{equation}%
is the derivative of the zeta function. See the derivation in Methods \ref%
{dest}. On the other hand, the critical temperature of the solution of Eq.(%
\ref{GapEqB}) is almost irrelevant to $J$.

This peculiar behavior of the critical temperature (\ref{Tor}) is
intuitively understood as follows. The singlet superconducting gap and the
out-of-plane triplet superconducting gap are formed by the Cooper pair of
electron with opposite spins. The electrons with $\mathbf{k}$ and $-\mathbf{k%
}$ have the same spin in the case of the altermagnet, while the electrons
with $\mathbf{k}$ and $-\mathbf{k}$ have opposite spins in the case of
odd-parity magnets. It contradicts with the singlet superconducting gap and
the out-of-plane triplet superconducting gap. Hence, their critical
temperature is lowered. On the other hand, the in-plane triplet
superconducting gap is formed by the same spins. Hence, it does not
contradict with the altermagnetic spin splitting and their critical
temperature is scarcely affected, while it contradicts with the odd-parity
magnetic spin splitting and their critical temperature is lowered.

It is also mathematically understood as follows. Eq.(\ref{GapEqB}) contains
the BCS kernel 
\begin{equation}
K_{\text{BCS}}\left( \mathbf{k}\right) =\sum_{s=\pm 1}\frac{\tanh \frac{%
\varepsilon _{s}\left( \mathbf{k}\right) }{2k_{\text{B}}T}}{4k_{\text{B}%
}T\varepsilon _{s}\left( \mathbf{k}\right) }.  \label{tanhA}
\end{equation}%
It is identical to the essential term contained in the well-known BCS
equation in the absence of the magnetic order, where the zero of the
denominator and the zero of the eliminator perfectly coincide. We show $K_{%
\text{BCS}}\left( \mathbf{k}\right) $\ in Fig.\ref{FigDensity}(a), where all
of the Fermi sea contribute to the superconductivity. Hence, the critical
temperature in the presence of magnetic order is almost identical to that in
the absence of the magnetic order. On the other hand, Eq.(\ref{GapEqA})
contains the term%
\begin{equation}
K_{\text{BCS}}^{\text{no}}\left( \mathbf{k}\right) =\sum_{s=\pm 1}\frac{%
\tanh \frac{\varepsilon _{s}\left( \mathbf{k}\right) }{2k_{\text{B}}T}}{4k_{%
\text{B}}TH_{\text{kine}}\left( \mathbf{k}\right) }.
\end{equation}%
Its integration over $\mathbf{k}$ is much smaller than the integration of
Eq.(\ref{tanhA}) over $\mathbf{k}$. We show $K_{\text{BCS}}^{\text{no}%
}\left( \mathbf{k}\right) $\ in Fig.\ref{FigDensity}(b), where only the
crossing points of the up and down spins have large values. It leads to the
lower critical temperature.

Indeed, we will see in the next sections that the $s$-wave and the extended $%
s$-wave and $d$-wave superconductivities are suppressed for large $J$ for
altermagnets, while it is not for odd-parity magnets. On the other hand, the
in-plane $p$-wave supercondutivity is not suppressed by the increase of $J$.
Hence, the in-plane $p$-wave supercondutor is realized for sufficiently
large $J$ for altermagnets.

\begin{figure}[t]
\centerline{\includegraphics[width=0.48\textwidth]{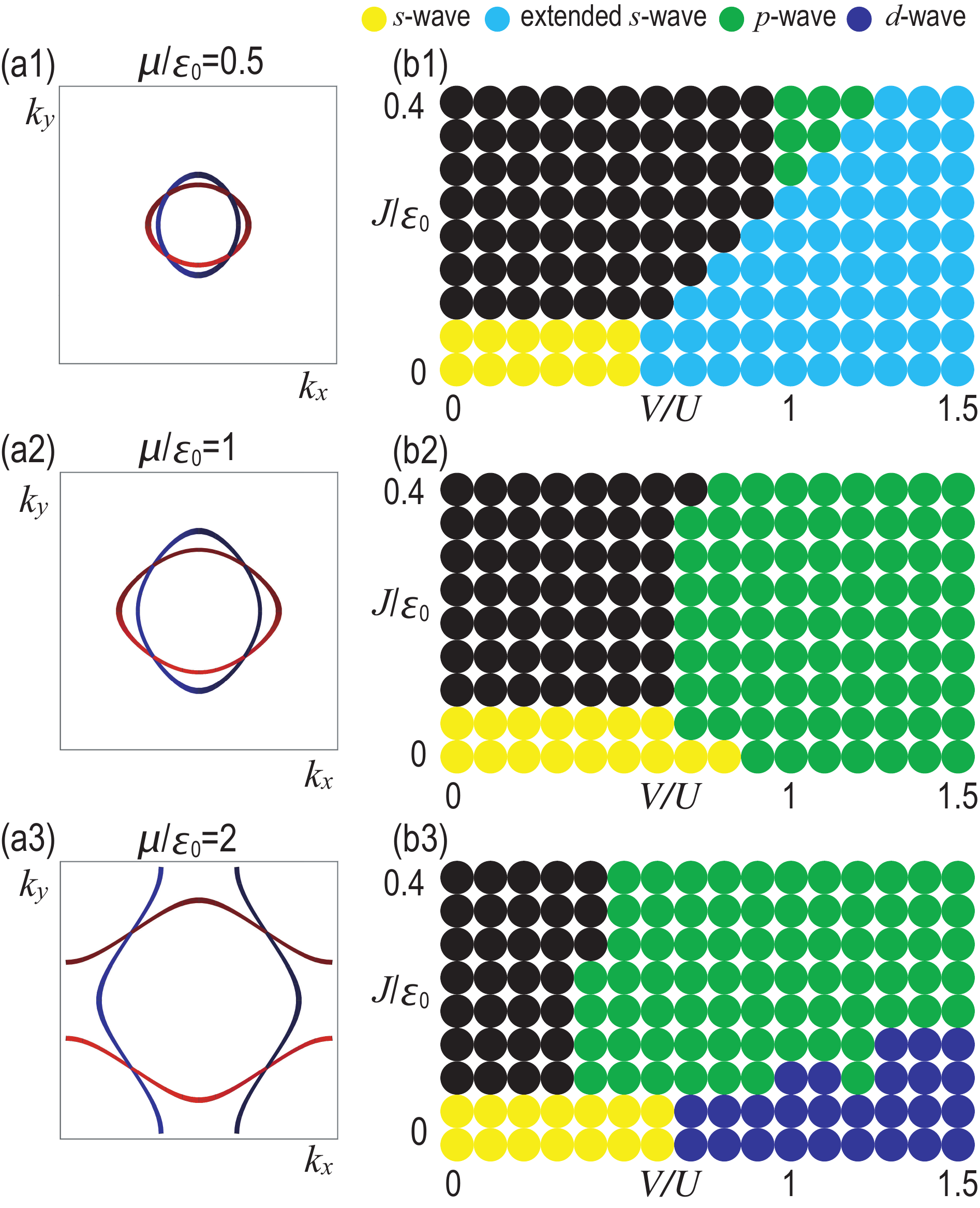}}
\caption{$d$-wave altermagnets. (a1), (a2) and (a3) Fermi surface. (b1),
(b2) and (b3) Phase diagram. (a1) and (b1) $\protect\mu /\protect\varepsilon %
_{0}=0.5$, (a2) and (b2) $\protect\mu /\protect\varepsilon _{0}=1$ and (a3)
and (b3) $\protect\mu /\protect\varepsilon _{0}=2$. The horizontal axis is $%
V/U$, while the vertical axis is $J/\protect\varepsilon _{0}$. Black disks
represent the absence of superconductivity. We have set $J/\protect%
\varepsilon _{0}=0.4$ for the Fermi surface. See also the caption of Fig.%
\protect\ref{FigMu} for other colors.}
\label{FigdSC}
\end{figure}
\begin{figure}[t]
\centerline{\includegraphics[width=0.48\textwidth]{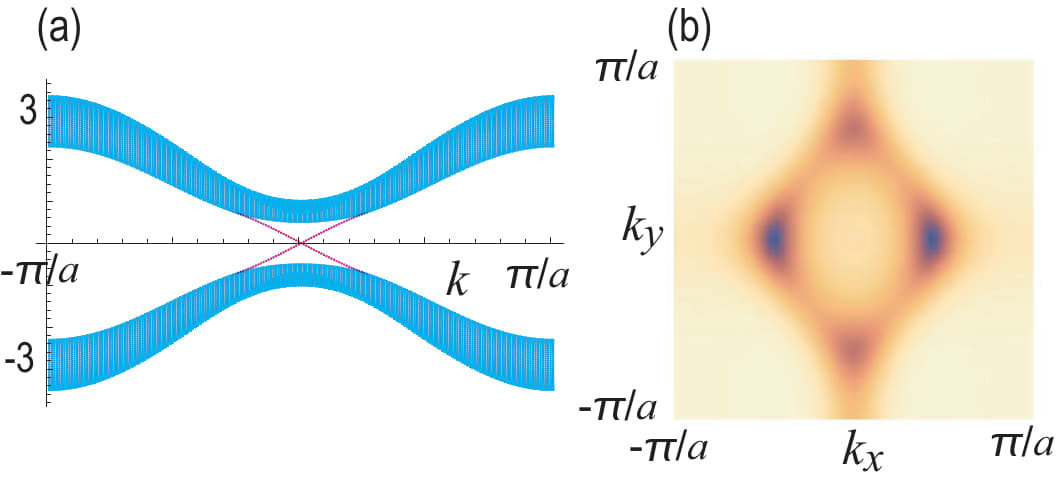}}
\caption{$d$-wave altermagnets. (a) Energy spectrum with the $p_{x}+ip_{y}$
chiral superconducting gap in nanoribbon geometry for $d$-wave altermagnets.
Red color indicates that the amplitude at the edges is large, while cyan
color indicates that it is small. We have set $t=-\protect\varepsilon _{0}/2,%
\protect\mu =\protect\varepsilon _{0}/2,J=0.2\protect\varepsilon _{0},\Delta
_{d}=\protect\varepsilon _{0}/2$ and $L=40$. (b) Momentum dependence of the
Berry curvature.}
\label{FigdRibbon}
\end{figure}

\subsubsection{$d$-wave altermagnets}

The phase diagram in the ($V/U,J$) plane is shown in Fig.\ref{FigdSC}.
First, we study the low chemical potential regime by setting $\mu
/\varepsilon _{0}=0.5$. The $s$-wave conductivity is destroyed for strong $J$%
\ because the spin singlet pairing is not compatible with the $d$-wave
altermagnetic Fermi surface. The chiral $p$-wave superconductor is realized
in the region around $V/U\simeq 1$\ and $J/\varepsilon _{0}\simeq 0.4$.
Next, we study the middle chemical potential regime ($\mu /\varepsilon
_{0}=1 $). The region of the $p$-wave superconductivity becomes broader
around $V/U\lesssim 1$\ for strong $J$\ because the $s$-wave
superconductivity is destabilized coupled with the $d$-wave altermagnetic
order. Finally, we study the phase diagram at the half filling ($\mu
/\varepsilon _{0}=2$). The $p$-wave superconductivity is stabilized instead
of the $d$-wave superconductivity for strong $J$. The phase diagram is
consistent with the previous result\cite{Hong}.

The band structure in nanoribbon geometry is shown in Fig.\ref{FigdRibbon}.
There are chiral edge states indicating the Majorana flat bands.
Correspondingly, the Chern number is evaluated to be 1 based on the
numerical integration of Eq.(\ref{Chern}) in Methods.

\begin{figure}[t]
\centerline{\includegraphics[width=0.48\textwidth]{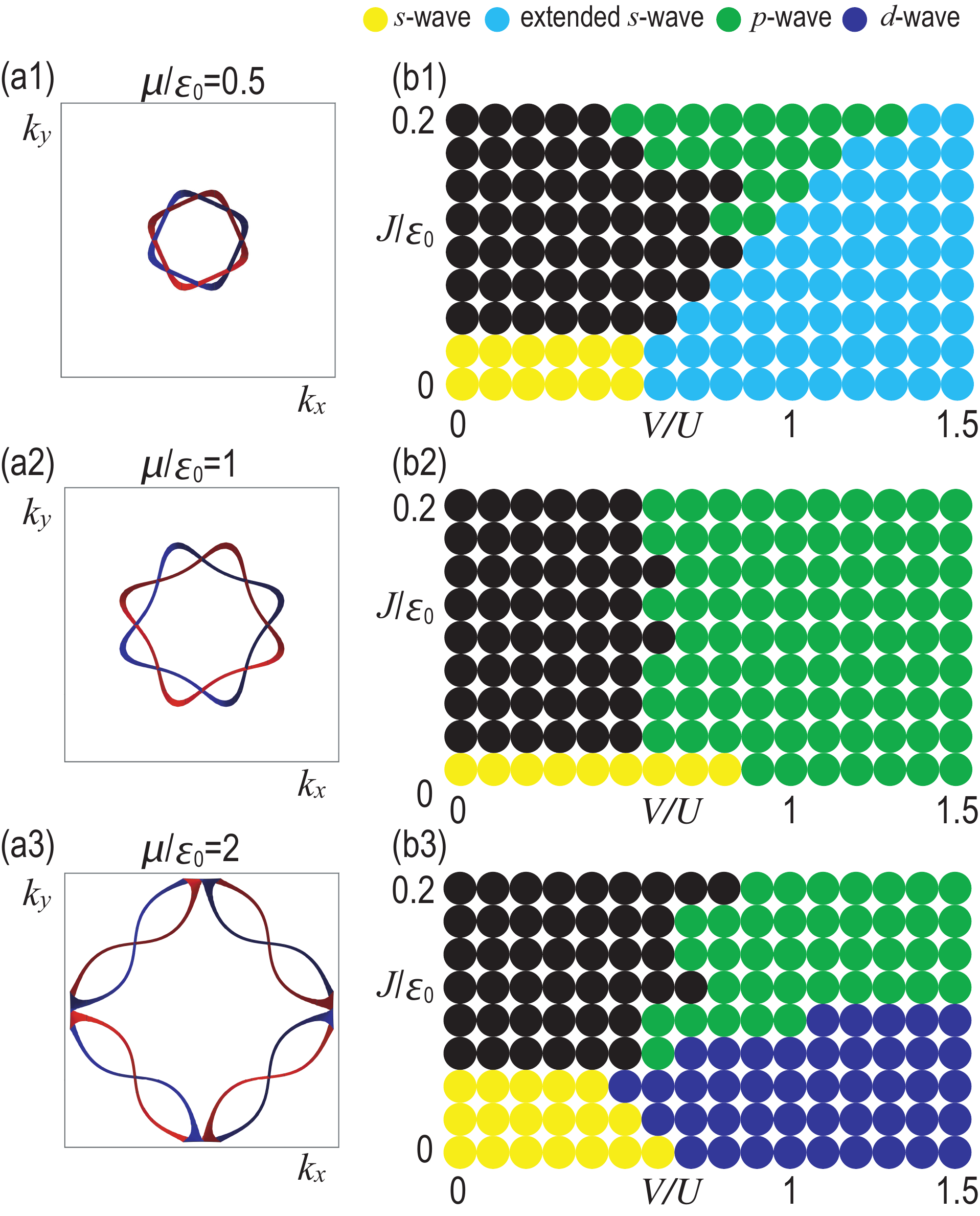}}
\caption{$g$-wave altermagnets. (a1)$\sim $(a3) Illustration of the Fermi
surface. (b1)$\sim $(b3) Phase diagram. We have set $J/\protect\varepsilon %
_{0}=0.1$ for the Fermi surface. See also the caption of Fig.\protect\ref%
{FigdSC}. }
\label{FiggSC}
\end{figure}
\begin{figure}[t]
\centerline{\includegraphics[width=0.48\textwidth]{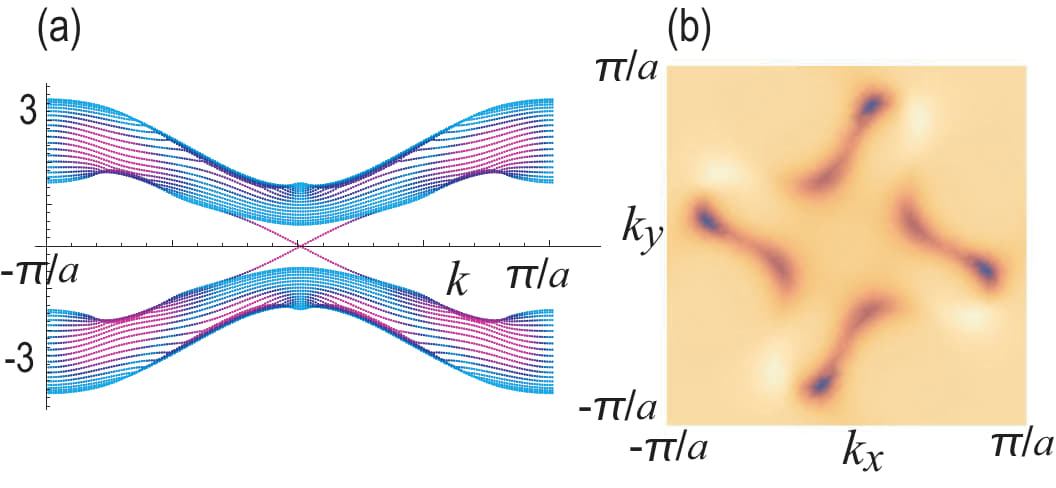}}
\caption{$g$-wave altermagnets. (a) Energy spectrum with the $p_{x}+ip_{y}$
chiral superconducting gap in nanoribbon geometry for $g$-wave altermagnets.
We have set $t=-\protect\varepsilon _{0}/2,\protect\mu =\protect\varepsilon %
_{0}/2,J=0.2\protect\varepsilon _{0},\Delta _{g}=\protect\varepsilon _{0}/2$
and $L=40$. (b) Momentum dependence of the Berry curvature.}
\label{FiggRibbon}
\end{figure}

\subsubsection{$g$-wave altermagnets}

The phase diagram in the ($V/U,J$) plane is shown in Fig.\ref{FiggSC}. The
feature of the phase diagram is almost identical to that of the $d$-wave
altermagnet although the positions of phase boundaries are different.

The band structure in nanoribbon geometry is shown in Fig.\ref{FiggRibbon}.
There are chiral edge states indicating the Majorana flat bands.
Correspondingly, the Chern number is evaluated to be 1 based on the
numerical integration of Eq.(\ref{Chern}) in Methods.

\subsubsection{$i$-wave altermagnets}

\begin{figure}[t]
\centerline{\includegraphics[width=0.48\textwidth]{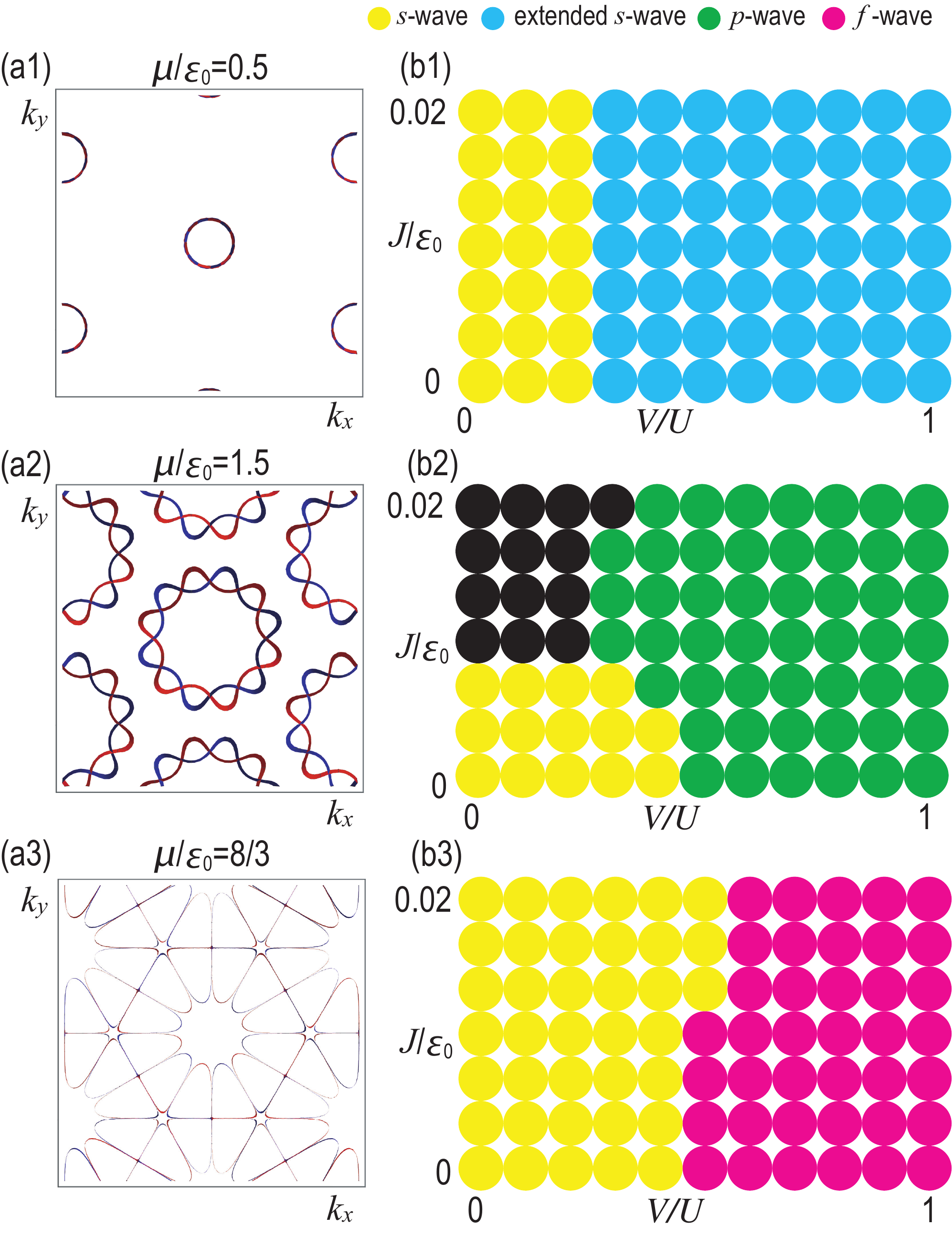}}
\caption{$i$-wave altermagnets. (a1)$\sim $(a3) Illustration of the Fermi
surface. (b1)$\sim $(b3) Phase diagram. (a1) and (b1) $\protect\mu /\protect%
\varepsilon _{0}=0.5$, (b1) and (b2) $\protect\mu /\protect\varepsilon %
_{0}=1.5$, (a3) and (b3) $\protect\mu /\protect\varepsilon _{0}=8/3$. We
have set $J/\protect\varepsilon _{0}=0.025$ for the Fermi surface. }
\label{FigiSC}
\end{figure}

The phase diagram in the ($V/U,J$) plane is shown in Fig.\ref{FigiSC}. The
phase diagrams for the low and middle chemical potential regimes are similar
to those of the $d$-wave and $g$-wave altermagnets. We study the chemical
potential at the van Hove singularity ($\mu /\varepsilon _{0}=8/3$).

\begin{figure}[t]
\centerline{\includegraphics[width=0.48\textwidth]{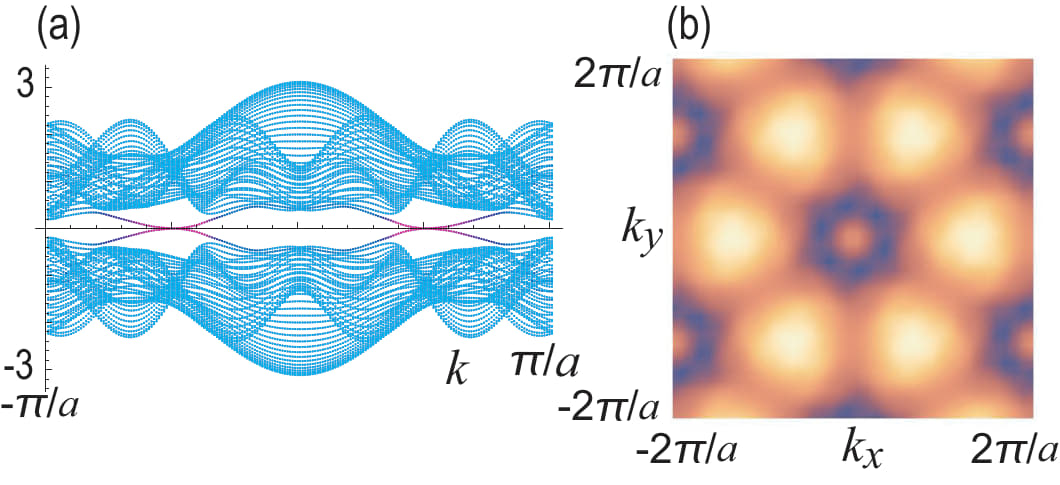}}
\caption{$i$-wave altermagnets. (a) Energy spectrum with the $p_{x}+ip_{y}$
chiral superconducting gap in nanoribbon geometry for $i$-wave altermagnets.
We have set $t=-2\protect\varepsilon _{0}/3,\protect\mu =2\protect%
\varepsilon _{0},J=0.02\protect\varepsilon _{0},\Delta _{i}=\protect%
\varepsilon _{0}/2$ and $L=40$. (b) Momentum dependence of the Berry
curvature. }
\label{FigiRibbon}
\end{figure}

The band structure in nanoribbon geometry is shown in Fig.\ref{FigiRibbon}.
There are two chiral edge states indicating the Majorana zero modes. The
Chern number is evaluated to be 2 based on the numerical integration
of Eq.(\ref{Chern}) in Methods.

\subsection{Odd-parity magnets}

Odd-parity magnets preserve time-reversal symmetry. It leads to the class
DIII topological superconductors, where $\mathbb{Z}_{2}$ index is the
topological number. In the case of the $X$-wave odd-parity magnet with $X=p$
and $f$, $X$-wave superconductor has point nodes with the 2 nodes and 6
nodes for $X=p$ and $f$, respectively. We show that Majorana flat bands
emerge for the momentum $k_{y}$ between the point nodes. They are
characterized by nontrivial winding numbers $W\left( k_{y}\right) $.

\subsubsection{$p$-wave magnets}

\begin{figure}[t]
\centerline{\includegraphics[width=0.48\textwidth]{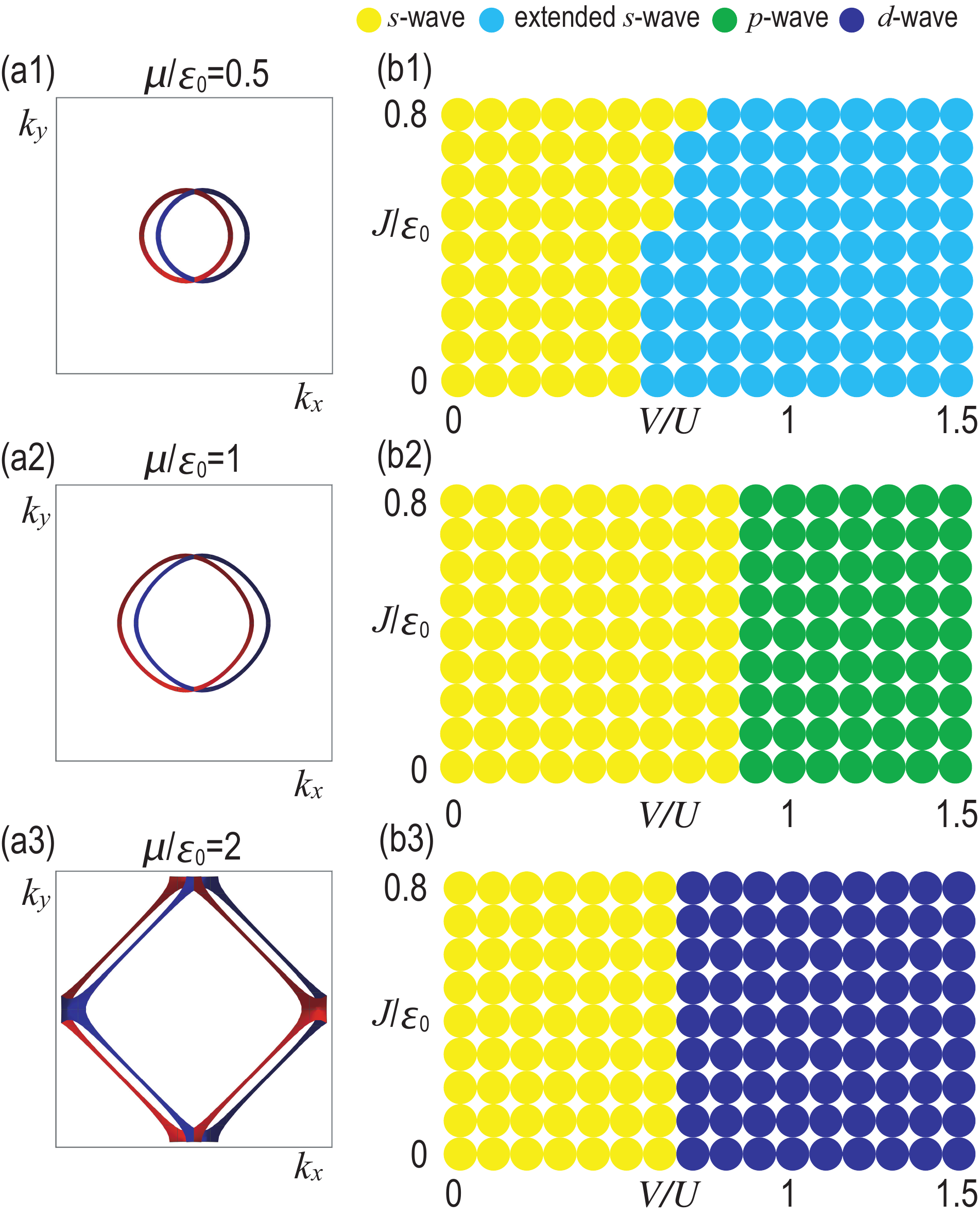}}
\caption{$p$-wave magnets. (a1)$\sim $(a3) Illustration of the Fermi
surface. (b1)$\sim $(b3) Phase diagram. See also the caption of Fig.\protect
\ref{FigdSC}. We have set $J/\protect\varepsilon _{0}=0.2$ for the Fermi
surface. }
\label{FigpSC}
\end{figure}

The phase diagram in the ($V/U,J$) plane is shown in Fig.\ref{FigpSC}. The
phase boundary is almost independent of $J$, which is highly contrasted with
that of altermagnets. It is because the $s$-wave superconductivity is not
destroyed by the presence of the $p$-wave magnetic order.

\begin{figure}[t]
\centerline{\includegraphics[width=0.48\textwidth]{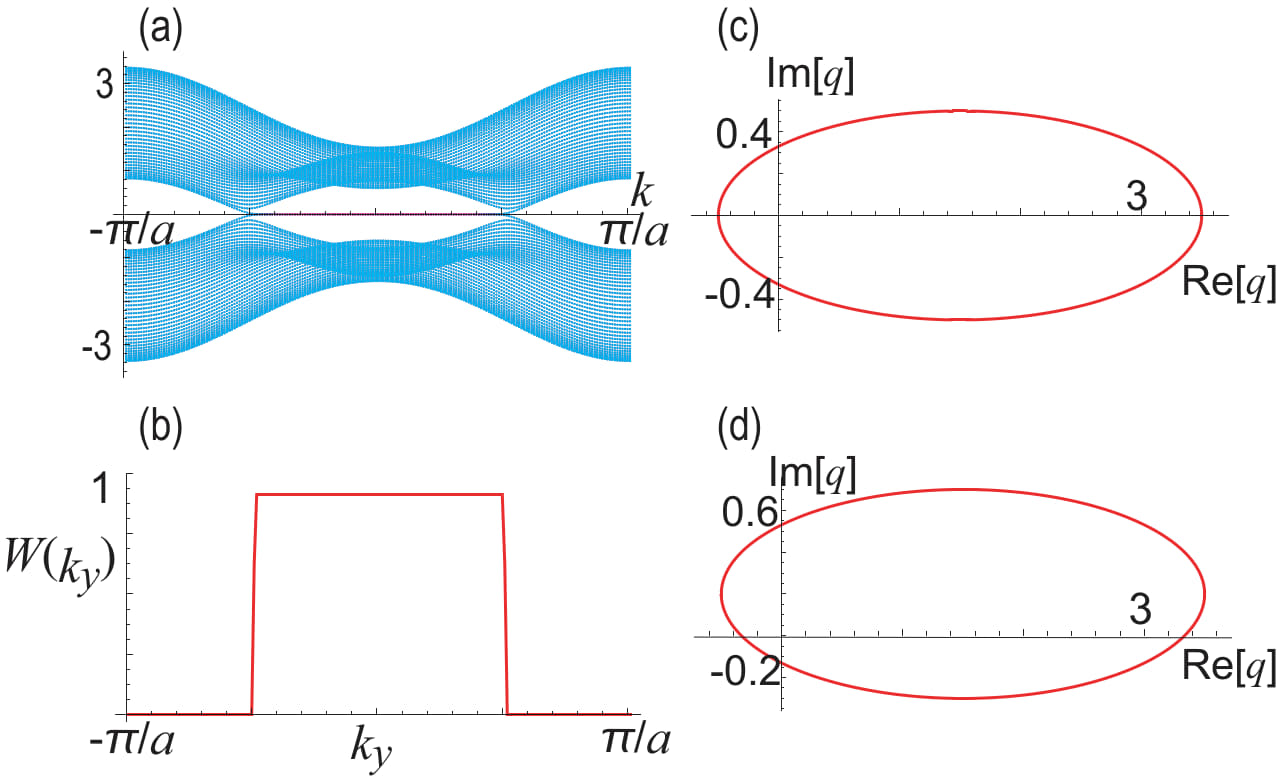}}
\caption{$p$-wave magnets. (a) Energy spectrum in nanoribbon geometry
for $p$-wave magnets with $\left( p_{x}+s\right) $-wave superconductivity.
(b) Winding number as a function of $k_{y}$, which corresponds to the
emergence of Majorana flat bands in (a). (c) Polar plot of $q\left( \mathbf{k%
}\right) $ at $k_{y}=0$ in the complex plane for the $p_{x}$-wave
superconductivity. (d) Polar plot of $q\left( \mathbf{k}\right) $ at $%
k_{y}=0 $ in the complex plane for the $\left( p_{x}+s\right) $-wave
superconductivity. We have set $t=-\protect\varepsilon _{0}/2,\protect\mu =2%
\protect\varepsilon _{0},J=0.2\protect\varepsilon _{0},\Delta _{p}=\protect%
\varepsilon _{0}/2,\Delta _{s}=0.1\protect\varepsilon _{0}$ and $L=40$.}
\label{FigpRibbon}
\end{figure}

The band structure of the $p_{x}$-wave superconductor in nanoribbon geometry
is shown in Fig.\ref{FigpRibbon}(a). There are flat-band zero-energy edge
states indicating the Majorana flat bands. The emergence of the Majorana
flat bands well agrees with the winding number as shown in Fig.\ref%
{FigpRibbon}(b). The momentum $k_{y}$ dependent winding number is defined by%
\cite{Ryu,KYKim} 
\begin{equation}
W\left( k_{y}\right) \equiv \frac{1}{2\pi }\int_{0}^{2\pi }\frac{1}{q(%
\mathbf{k})}\frac{\partial q(\mathbf{k})}{\partial k_{x}}dk_{x}
\label{W}
\end{equation}%
with 
\begin{equation}
q\left( \mathbf{k}\right) =2t\left( \cos ak_{x}+\cos ak_{y}-\mu \right)
+J\sin ak_{x}+i\Delta _{p}\sin ak_{x}.
\end{equation}%
See the details in Methods \ref{Wind}. It is explicitly evaluated as $%
W\left( k_{y}\right) =1$ for $\left\vert \mu -\cos ak_{y}\right\vert <1$ and 
$W\left( k_{y}\right) =0$ for $\left\vert \mu -\cos ak_{y}\right\vert >1$.
The Majorana flat bands appear for the momentum satisfying $W\left(
k_{y}\right) =1$.

Actually, there is a mixing of the $s$-wave and $p_{x}$-wave
superconductivities\cite{Kho,KYKim,Carmero} due to the off-diagonal term in
Eq.(\ref{LGB}) in Methods. Even in this case, Majorana flat bands remain as
they are although the band crossing momenta change. This is\ understood in
terms of the winding number (\ref{W}) of $q_{p}\left( \mathbf{k}\right) $
defined by%
\begin{align}
q_{p}\left( \mathbf{k}\right) =& 2t\left( \cos ak_{x}+\cos ak_{y}-\mu
\right) +J\sin ak_{x}  \notag \\
& +i\left( \Delta _{s}+\Delta _{p}\sin ak_{x}\right) .  \label{qp}
\end{align}%
The winding number is%
\begin{equation}
W\left( k_{y}\right) =1
\end{equation}%
for%
\begin{equation}
\left( \frac{J\Delta _{s}}{2t\Delta _{p}}+\mu -\cos ak_{y}\right)
^{2}+\left( \frac{\Delta _{s}}{\Delta _{p}}\right) ^{2}<1,
\end{equation}%
while 
\begin{equation}
W\left( k_{y}\right) =0
\end{equation}%
for%
\begin{equation}
\left( \frac{J\Delta _{s}}{2t\Delta _{p}}+\mu -\cos ak_{y}\right)
^{2}+\left( \frac{\Delta _{s}}{\Delta _{p}}\right) ^{2}>1.
\end{equation}%
See the derivation in Methods \ref{Wind}.

\subsubsection{$f$-wave magnets}

\begin{figure}[t]
\centerline{\includegraphics[width=0.48\textwidth]{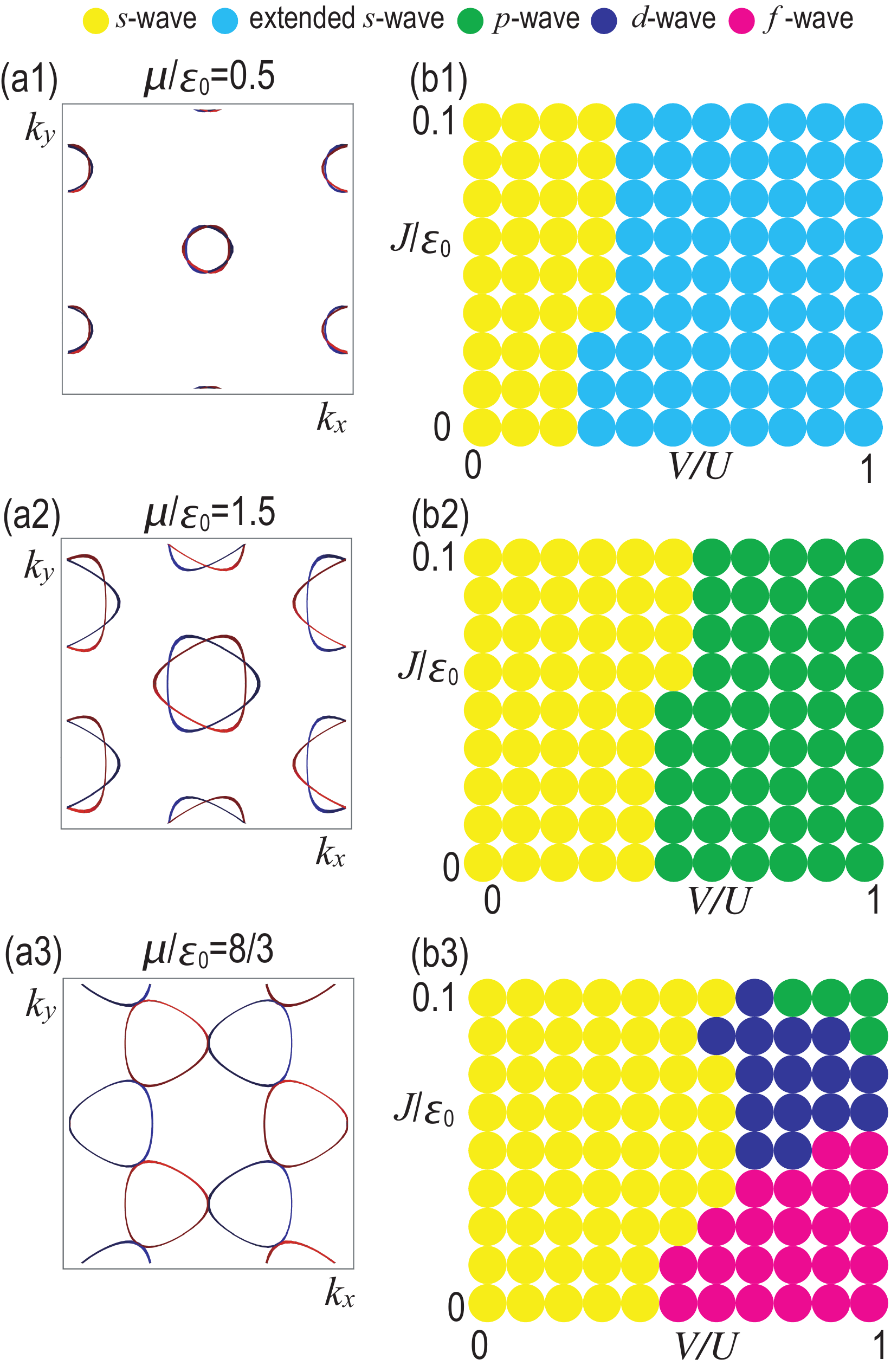}}
\caption{$f$-wave magnets. Illustration of the Fermi surface. We have set $J/%
\protect\varepsilon _{0}=0.05$ for the Fermi surface. See also the caption
of Fig.\protect\ref{FigiSC}.}
\label{FigfSC}
\end{figure}

The phase diagram in the ($V/U,J$) plane is shown in Fig.\ref{FigfSC}. The
phase diagram of the $f$-wave magnet is similar to that of the $p$-wave
magnet.

\begin{figure}[t]
\centerline{\includegraphics[width=0.48\textwidth]{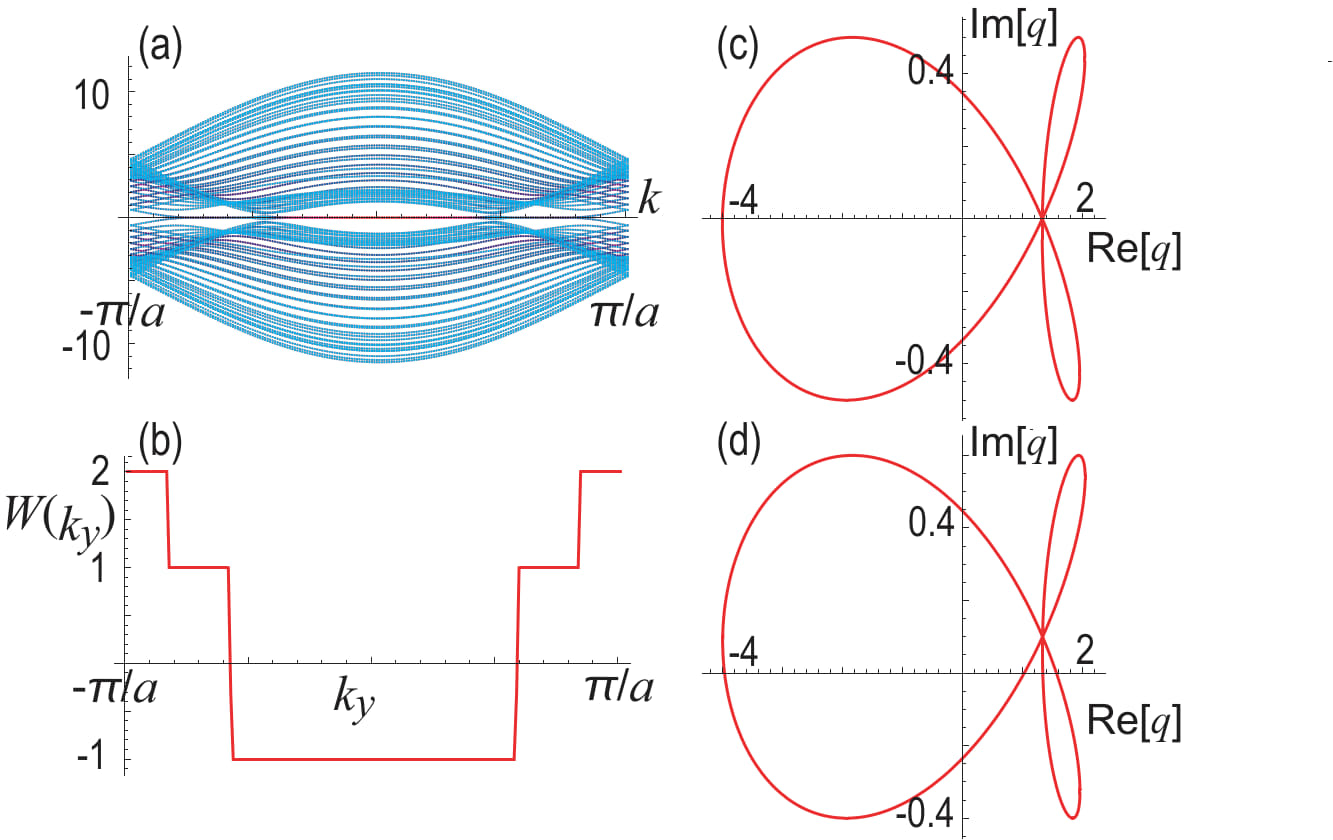}}
\caption{$f$-wave magnets. Energy spectrum in nanoribbon geometry for $%
f $-wave magnets with $\left( f+s\right) $-wave superconductivity. We have
set $t=-2\protect\varepsilon _{0}/3,\protect\mu =8\protect\varepsilon %
_{0}/3,J=0.05\protect\varepsilon _{0}$, $\Delta _{f}=\protect\varepsilon %
_{0}/2,\Delta _{s}=0.1\protect\varepsilon _{0}$ and $L=40$. See also the
caption of Fig.\protect\ref{FigpRibbon}.}
\label{FigfRibbon}
\end{figure}

The band structure of the $f$-wave superconductivity with ($p_{x}+s$)
superconductivity in nanoribbon geometry is shown in Fig.\ref{FigfRibbon}%
(a). There are flat-band zero-energy edge states indicating the Majorana
falt bands. The winding number (\ref{W}) is shown in Fig.\ref{FigfRibbon}%
(b). The winding number takes $W\left( k_{y}\right) =2,1$, and $-1$.\
However, the topological number should be $\mathbb{Z}_{2}$ because the
topological class is DIII. Hence, we use mod$_{2}W\left( k_{y}\right) $ as
the topological number. Correspondingly, there are no Majorana flat bands
around $k_{y}=\pi /a$, where $W\left( k_{y}\right) =2$, while there emerge
Majorana flat bands for $W\left( k_{y}\right) =\pm 1$. We also show $%
q_{f}\left( \mathbf{k}\right) $,%
\begin{align}
q_{f}\left( \mathbf{k}\right) =& 2t\left( \cos ak_{x}+\cos a\frac{k_{x}-%
\sqrt{3}k_{y}}{2}+\cos a\frac{k_{x}+\sqrt{3}k_{y}}{2}\right)  \notag \\
& -6t-\mu +Jf_{f}\left( \mathbf{k}\right) +i\left( \Delta _{s}+\Delta
_{f}f_{f}\left( \mathbf{k}\right) \right)
\end{align}%
with%
\begin{equation}
f_{f}\left( \mathbf{k}\right) \equiv \frac{16}{3\sqrt{3}}\left( \sin
ak_{x}\sin a\frac{k_{x}+\sqrt{3}k_{y}}{2}\sin a\frac{k_{x}-\sqrt{3}k_{y}}{2}%
\right)
\end{equation}%
in the complex plane in Fig.\ref{FigfRibbon}(c) and (d) for the $p$-wave
superconductivity and ($p_{x}+s$) superconductivity, respectively. The
topological superconductivity is robust against the mixing of the $s$-wave
superconductivity because the effect of the $s$-wave superconductivity is
the shift of $q\left( \mathbf{k}\right) $ along the imaginary direction.
Indeed, the Majorana flat bands are robust in the presence of the mixing of
the $s$-wave superconductivity as shown in Fig.\ref{FigfRibbon}(a).

\section{Discussion}

We have systematically studied the phase diagram of superconductors coupled
with the $X$-wave magnet. The phase diagram is essentially different between
the altermagnets and the odd-parity magnets. The main reason is that the $s$%
-wave singlet paring is not compatible with the altermagnetic order but is
compatible with the odd-parity magnetic order. We find class D chiral $%
p_{x}+ip_{y}$-wave topological superconductors are realized in altermagnets
accompanied with chiral edge modes, while class DIII $p+s$ ($f+s$)
topological superconductors are realized in $p$ ($f$)-wave magnets
accompanied with Majorana flat bands between the superconducting gap nodes.

\section{Methods}

\subsection{Linearized gap equations}

\label{GapEq}

Superconducting gaps are summarized in terms of the $\mathbf{d}\left( 
\mathbf{k}\right) $ vector%
\begin{equation}
\Delta \left( \mathbf{k}\right) =i\left( \mathbf{d}\left( \mathbf{k}\right)
\cdot \mathbf{\sigma }\right) \sigma _{y}.
\end{equation}%
We expand the superconducting gap functions by the the orthogonal functions $%
\Gamma \left( \mathbf{k}\right) $ as%
\begin{equation}
\Delta _{m}\left( \mathbf{k}\right) =\Delta _{m}\Gamma _{m}\left( \mathbf{k}%
\right) .
\end{equation}%
We also expand the interaction by the function $\Gamma \left( \mathbf{k}%
\right) $ as\cite{Xiao} 
\begin{equation}
V_{m\ell }=\frac{1}{N^{2}}\sum_{\mathbf{k},\mathbf{k}^{\prime }}\text{Tr}\!%
\left[ \Gamma _{m}^{\dagger }(\mathbf{k})\,\hat{V}(\mathbf{k},\mathbf{k}%
^{\prime })\,\Gamma _{\ell }(\mathbf{k}^{\prime })\right] ,
\end{equation}%
where 
\begin{equation}
V(\mathbf{k},\mathbf{k}^{\prime })=U+2V\bigl[\cos (k_{x}-k_{x}^{\prime
})+\cos (k_{y}-k_{y}^{\prime })\bigr]
\end{equation}%
for the square lattice\cite{KYKim,Chak} and 
\begin{align}
&V(\mathbf{k},\mathbf{k}^{\prime })  \notag \\
=&U+2V\bigl[\cos (k_{x}-k_{x}^{\prime })+2\cos \frac{k_{x}-k_{x}^{\prime }}{2%
}\cos (\frac{\sqrt{3}}{2}\left( k_{y}-k_{y}^{\prime }\right) )\bigr]
\end{align}%
for the triangular lattice.

\subsubsection{Altermagnets}

Linearized gap equations for altermagnets are given by\cite{Xiao,Hong}%
\begin{equation}
\Lambda _{m^{\prime }l^{\prime }}(T)=-\frac{V_{m^{\prime }m^{\prime }}\,k_{%
\text{B}}T}{N_{\Gamma }}\sum_{\mathbf{k}}M_{m^{\prime }\ell ^{\prime }}^{%
\text{even}}  \label{LGA}
\end{equation}%
with 
\begin{align}
& M_{m^{\prime }\ell ^{\prime }}^{\text{even}}  \notag \\
\equiv & \left( 
\begin{array}{cccc}
\Gamma _{0}^{\dagger }\Gamma _{0}K_{\text{BCS}}^{\text{no}} & 0 & 0 & 0 \\ 
0 & \Gamma _{x}^{\dagger }\Gamma _{x}K_{\text{BCS}} & -i\Gamma _{x}^{\dagger
}\Gamma _{y}K_{\text{BCS}}^{s} & 0 \\ 
0 & i\Gamma _{y}^{\dagger }\Gamma _{x}K_{\text{BCS}}^{s} & \Gamma
_{y}^{\dagger }\Gamma _{y}K_{\text{BCS}} & 0 \\ 
0 & 0 & 0 & \Gamma _{z}^{\dagger }\Gamma _{z}K_{\text{BCS}}^{\text{no}}%
\end{array}%
\right) ,
\end{align}%
where%
\begin{equation}
K_{\text{BCS}}\left( \mathbf{k}\right) \equiv \sum_{s=\pm 1}\frac{\tanh 
\frac{\varepsilon _{s}\left( \mathbf{k}\right) }{2k_{\text{B}}T}}{4k_{\text{B%
}}T\varepsilon _{s}\left( \mathbf{k}\right) },
\end{equation}%
\begin{equation}
K_{\text{BCS}}^{s}\left( \mathbf{k}\right) \equiv \sum_{s=\pm 1}s\frac{\tanh 
\frac{\varepsilon _{s}\left( \mathbf{k}\right) }{2k_{\text{B}}T}}{4k_{\text{B%
}}T\varepsilon _{s}\left( \mathbf{k}\right) },
\end{equation}%
and%
\begin{equation}
K_{\text{BCS}}^{\text{no}}\equiv \sum_{s=\pm 1}\frac{\tanh \frac{\varepsilon
_{s}\left( \mathbf{k}\right) }{2k_{\text{B}}T}}{4k_{\text{B}}TH_{\text{kine}%
}\left( \mathbf{k}\right) }.
\end{equation}

\subsubsection{Odd-parity magnets}

Linearized gap equations for odd-parity magnets are given by\cite{KYKim}%
\begin{equation}
\Lambda _{m^{\prime }l^{\prime }}(T)=-\frac{V_{m^{\prime }m^{\prime }}\,k_{%
\text{B}}T}{N_{\Gamma }}\sum_{\mathbf{k}}M_{m^{\prime }\ell ^{\prime }}^{%
\text{odd}}  \label{LGB}
\end{equation}%
with%
\begin{equation}
M_{m^{\prime }\ell ^{\prime }}^{\text{odd}}\equiv \left( 
\begin{array}{cccc}
\Gamma _{0}^{\dagger }\Gamma _{0}K_{\text{BCS}} & 0 & 0 & \Gamma
_{0}^{\dagger }\Gamma _{z}K_{\text{BCS}}^{s} \\ 
0 & \Gamma _{x}^{\dagger }\Gamma _{x}K_{\text{BCS}}^{\text{no}} & 0 & 0 \\ 
0 & 0 & \Gamma _{y}^{\dagger }\Gamma _{y}K_{\text{BCS}}^{\text{no}} & 0 \\ 
\Gamma _{z}^{\dagger }\Gamma _{0}K_{\text{BCS}}^{s} & 0 & 0 & \Gamma
_{z}^{\dagger }\Gamma _{z}K_{\text{BCS}}%
\end{array}%
\right) .
\end{equation}%
There are off-diagonal components, which leads to the mixing of spin-singlet
and spin-triplet superconductivities\cite{KYKim,Carmero}. However, numerical
estimations show that these mixing is tiny for $J\lesssim t$. Hence, it is
reasonable to assume that there is no mixing for $J\lesssim t$.

\subsection{Superconducting paring functions}

The on-site attractive interaction $U$ in Eq.(\ref{Hubbard}) induces the
spin singlet $s$-wave superconducting gap without the $\mathbf{k}$\
dependence%
\begin{equation}
\Delta _{s}\left( \mathbf{k}\right) =\Delta _{s},
\end{equation}%
while the nearest-neighbor attractive interaction $V$ in Eq.(\ref{Hubbard})
may induce the spin-singlet extended $s$-wave superconducting gap%
\begin{equation}
\Delta _{\text{ex-}s}\left( \mathbf{k}\right) =\Delta _{\text{ex-}s}\left(
\cos ak_{x}+\cos ak_{y}\right) ,  \label{exs}
\end{equation}%
or the spin-singlet $d$-wave superconducting gap%
\begin{equation}
\Delta _{d}\left( \mathbf{k}\right) =\Delta _{d}\left( \cos ak_{x}-\cos
ak_{y}\right) ,
\end{equation}%
or the spin-triplet chiral $p$-wave superconducting gap%
\begin{equation}
\Delta _{p}\left( \mathbf{k}\right) =\Delta _{p}\left( \sin ak_{x}+i\sin
ak_{y}\right) .
\end{equation}%
The spin-triplet chiral $p$-wave superconductor is a topological
superconductor, which hosts Majorana chiral-edge states.

We also study the superconductivity on the triangular lattice. The on-site
attractive interaction $U$ induces the spin singlet $s$-wave superconducting
gap%
\begin{equation}
\Delta _{s}\left( \mathbf{k}\right) =\Delta _{s},
\end{equation}%
while the nearest-neighbor attractive interaction $V$ may induce the
spin-singlet extended $s$-wave superconducting gap%
\begin{equation}
\Delta _{\text{ex-}s}\left( \mathbf{k}\right) =\Delta _{\text{ex-}%
s}\sum_{j}\cos \left( \mathbf{n}_{j}\cdot a\mathbf{k}\right)
\end{equation}%
with%
\begin{equation}
\mathbf{n}_{j}=\left( \cos \frac{2\pi j}{3},\sin \frac{2\pi j}{3}\right) ,
\label{nj}
\end{equation}%
or the spin-triplet chiral $p$-wave superconducting gap%
\begin{equation}
\Delta _{p}\left( \mathbf{k}\right) =\Delta _{p}\sum_{j}\omega ^{j}\sin
\left( \mathbf{n}_{j}\cdot a\mathbf{k}\right)
\end{equation}%
with $\omega \equiv e^{2\pi i/3}$ and Eq.(\ref{nj}), or the spin-singlet $d$%
-wave superconducting gap%
\begin{align}
\Delta _{d}\left( \mathbf{k}\right) =& \Delta _{d}\sum_{j}\cos ak_{x}-\cos 
\frac{ak_{x}}{2}\cos \frac{\sqrt{3}ak_{y}}{2}  \notag \\
& +i\sin \frac{ak_{x}}{2}\sin \frac{\sqrt{3}ak_{y}}{2},
\end{align}%
or the spin-singlet $f$-wave superconducting gap%
\begin{equation}
\Delta _{f}\left( \mathbf{k}\right) =\Delta _{f}\sum_{j}\sin \left( \mathbf{n%
}_{j}\cdot a\mathbf{k}\right)
\end{equation}%
with Eq.(\ref{nj}). Which superconductivity is realized should be determined
by numerically solving the linearized gap equation.

\subsection{Destabilization of superconductivity}

\label{dest}

We derive the critical temperature Eq.(\ref{Tor}) in the main text. The
linearized gap equation reads%
\begin{equation}
1=V\sum_{\mathbf{k}}\frac{\sum_{s=\pm 1}\tanh \frac{\xi \left( \mathbf{k}%
\right) +sJf\left( \mathbf{k}\right) }{2T}}{2\xi \left( \mathbf{k}\right) }.
\end{equation}%
We expand it with $Jf\left( \mathbf{k}\right) /2T_{\text{cr}}$,%
\begin{align}
1=& VN\left( 0\right) \int \frac{d\Omega _{\mathbf{k}}}{S_{\text{FS}}}%
\int_{0}^{\omega _{\text{D}}}\frac{d\xi }{\xi }  \notag \\
& \left[ 2\tanh \frac{\xi }{2T_{\text{cr}}}-\left( \frac{Jf\left( \mathbf{k}%
\right) }{2T_{\text{cr}}}\right) ^{2}2\tanh \frac{\xi }{2T_{\text{cr}}}\text{%
sech}^{2}\frac{\xi }{2T_{\text{cr}}}\right]  \notag \\
\simeq & VN\left( 0\right) \int_{0}^{\omega _{\text{D}}}\frac{d\xi }{\xi } 
\notag \\
& \left[ 2\tanh \frac{\xi }{2T_{\text{cr}}}-\frac{J^{2}\left\langle f\left( 
\mathbf{k}\right) ^{2}\right\rangle _{\text{FS}}}{2T_{\text{cr}}^{2}}\tanh 
\frac{\xi }{2T_{\text{cr}}}\text{sech}^{2}\frac{\xi }{2T_{\text{cr}}}\right]
,
\end{align}%
where we have used 
\begin{equation}
\tanh \left( x\pm \delta \right) =\tanh x\pm \delta \text{sech}^{2}x-\frac{%
\delta ^{2}}{2}2\tanh x\text{sech}^{2}x
\end{equation}%
and%
\begin{equation}
\sum_{\pm }\tanh \left( x\pm \delta \right) =2\tanh x-2\delta ^{2}\tanh x%
\text{sech}^{2}x.
\end{equation}%
We define%
\begin{equation}
F\left( T\right) \equiv UN\left( 0\right) \int_{0}^{\omega _{\text{D}}}\frac{%
d\xi }{\xi }\tanh \frac{\xi }{2T},
\end{equation}%
which satisfies the condition%
\begin{equation}
F\left( T_{\text{cr}}^{\left( 0\right) }\right) =1.
\end{equation}%
We expand it as%
\begin{equation}
F\left( T_{\text{cr}}\right) -1\simeq \left( T_{\text{cr}}-T_{\text{cr}%
}^{\left( 0\right) }\right) F^{\prime }\left( T_{\text{cr}}^{\left( 0\right)
}\right) =F^{\prime }\left( T_{\text{cr}}^{\left( 0\right) }\right) \delta T
\end{equation}%
with%
\begin{align}
& F^{\prime }\left( T_{\text{cr}}^{\left( 0\right) }\right) \delta T  \notag
\\
=& VN\left( 0\right) \left\langle f\left( \mathbf{k}\right)
^{2}\right\rangle _{\text{FS}}\int_{0}^{\omega _{\text{D}}}\frac{d\xi }{\xi }%
\frac{J^{2}}{2T_{\text{cr}}^{2}}\tanh \frac{\xi }{2T_{\text{cr}}}\text{sech}%
^{2}\frac{\xi }{2T_{\text{cr}}}  \notag \\
\simeq & VN\left( 0\right) \left\langle f\left( \mathbf{k}\right)
^{2}\right\rangle _{\text{FS}}\int_{0}^{\omega _{\text{D}}}\frac{d\xi }{\xi }%
\frac{J^{2}}{2T_{\text{cr}}^{2}}\tanh \frac{\xi }{2T_{\text{cr}}^{\left(
0\right) }}\text{sech}^{2}\frac{\xi }{2T_{\text{cr}}^{\left( 0\right) }},
\end{align}%
where%
\begin{equation}
\left\langle f\left( \mathbf{k}\right) ^{2}\right\rangle _{\text{FS}}\equiv
\int_{\text{FS}}\frac{d\mathbf{k}}{S_{\text{FS}}}f\left( \mathbf{k}\right)
^{2}
\end{equation}%
is the integration over the Fermi surface. Then, the $\delta T$ is obtained
as%
\begin{equation}
\delta T=\frac{VN\left( 0\right) \left\langle f\left( \mathbf{k}\right)
^{2}\right\rangle _{\text{FS}}\int_{0}^{\omega _{\text{D}}}\frac{d\xi }{\xi }%
\frac{J^{2}}{2T_{\text{cr}}^{2}}\tanh \frac{\xi }{2T_{\text{cr}}^{\left(
0\right) }}\text{sech}^{2}\frac{\xi }{2T_{\text{cr}}^{\left( 0\right) }}}{%
F^{\prime }\left( T_{\text{cr}}^{\left( 0\right) }\right) },
\end{equation}%
where%
\begin{align}
F^{\prime }\left( T\right) =& VN\left( 0\right) \int_{0}^{\omega _{\text{D}}}%
\frac{d\xi }{\xi }\left[ \frac{\partial }{\partial T}\tanh \frac{\xi }{2T}%
\right]  \notag \\
=& -VN\left( 0\right) \int_{0}^{\omega _{\text{D}}}d\xi \text{sech}^{2}\frac{%
\xi }{2T_{\text{cr}}^{\left( 0\right) }}d\xi .
\end{align}%
$\delta T$ is further calculated as%
\begin{equation}
\delta T\equiv T_{\text{cr}}-T_{\text{cr}}^{\left( 0\right) }=-C\left\langle
f\left( \mathbf{k}\right) ^{2}\right\rangle _{\text{FS}}\left( \frac{J}{T_{%
\text{cr}}^{\left( 0\right) }}\right) ^{2}.
\end{equation}%
$C$ is calculated as%
\begin{align}
C\equiv & \frac{\int_{0}^{\omega _{\text{D}}}\frac{\tanh \frac{\xi }{2T_{%
\text{cr}}^{\left( 0\right) }}\text{sech}^{2}\frac{\xi }{2T_{\text{cr}%
}^{\left( 0\right) }}}{\xi }d\xi }{-4\int_{0}^{\omega _{\text{D}}}\text{sech}%
^{2}\frac{\xi }{2T_{\text{cr}}^{\left( 0\right) }}d\xi } \\
\simeq & \frac{-28\zeta ^{\prime }\left( -2\right) }{-\frac{1}{T_{\text{cr}%
}^{\left( 0\right) }}}\simeq 28\zeta ^{\prime }\left( -2\right) T_{\text{cr}%
}^{\left( 0\right) },
\end{align}%
where we have used%
\begin{equation}
-4\int_{0}^{\omega _{\text{D}}}\text{sech}^{2}\frac{\xi }{2T_{\text{cr}%
}^{\left( 0\right) }}d\xi =-\frac{\tanh \frac{\omega _{\text{D}}}{2T_{\text{%
cr}}^{\left( 0\right) }}}{T_{\text{cr}}^{\left( 0\right) }}\simeq -\frac{1}{%
T_{\text{cr}}^{\left( 0\right) }}
\end{equation}%
for $\omega _{\text{D}}\gg 2T_{\text{cr}}^{\left( 0\right) }$, and%
\begin{equation}
\int_{0}^{\infty }\frac{\tanh \frac{\xi }{2}\text{sech}^{2}\frac{\xi }{2}}{%
\xi }=-28\zeta ^{\prime }\left( -2\right) ,
\end{equation}%
where%
\begin{equation}
\zeta ^{\prime }\left( x\right) \equiv \frac{d\zeta \left( x\right) }{dx}
\end{equation}%
is the derivative of the zeta function. Finally, we obtain the result%
\begin{equation}
\delta T\simeq 28\zeta ^{\prime }\left( -2\right) \frac{J^{2}\left\langle
f\left( \mathbf{k}\right) ^{2}\right\rangle _{\text{FS}}}{T_{\text{cr}%
}^{\left( 0\right) }},
\end{equation}%
which is Eq.(\ref{Tor}) in the main text.

\subsection{Tight-binding models}

\label{tight}

The tight-binding Hamiltonian is given by%
\begin{equation}
H=H_{\text{Kine}}+H_{X}.  \label{T-Hamil}
\end{equation}%
The kinetic term reads%
\begin{equation}
H_{\text{Kine,Sq}}=2t\left( 2-\cos ak_{x}-\cos ak_{y}\right)  \label{T-sq}
\end{equation}%
on the square lattice, while it reads%
\begin{equation}
H_{\text{Kine,Tri}}=2t\left( 3-\cos ak_{x}-\sum_{\pm }\cos a\frac{k_{x}\pm 
\sqrt{3}k_{y}}{2}\right)  \label{T-tri}
\end{equation}%
on the triangular lattice.

Tight-binding models for $p$-wave, $d$-wave, and $g$-wave altermagnets are
identical to those in the previous literature\cite%
{GI,Planar,MTJ,APEX,NLSeebeck}. The $X$-wave term $f_{X}$ is given by%
\begin{equation}
f_{p}=\sin ak_{x}
\end{equation}%
for the $p$-wave magnet\cite{pwave,EzawaPwave,EzawaPNeel,He,PEdel,Elliptic},
by%
\begin{equation}
f_{d}=2\left( \cos ak_{y}-\cos ak_{x}\right)
\end{equation}%
for the $d$-wave magnet\cite%
{SmejX,SmejX2,Zhu,Ghora,Li23,EzawaAlter,EzawaMetricC,EzawaVolta}, and by 
\begin{equation}
H_{g}=2\sin ak_{x}\sin ak_{y}\left( \cos ak_{y}-\cos ak_{z}\right)
\end{equation}%
for the $g$-wave magnet\cite{GI,Planar}.

On the other hand, we use similar tight-binding models for $f$-wave and $i$%
-wave magnets in the present manuscript comparing with those in the previous
literature\cite{GI,Planar,MTJ,APEX,NLSeebeck}.

\subsubsection{$f$-wave magnets}

We use the tight-binding model for the $f$-wave magnets,%
\begin{equation}
f_{f}=8\left( -\sin ak_{x}+\sin a\frac{k_{x}+\sqrt{3}k_{y}}{2}+\sin a\frac{%
k_{x}-\sqrt{3}k_{y}}{2}\right) .
\end{equation}%
It is constructed based on the nearest-neighbor sites. Note that it is
similar to the tight-binding model used in the previous literature\cite%
{GI,Planar,MTJ,APEX,NLSeebeck} but different, where the previous one is%
\begin{equation}
f_{f}=4\left( \sin ak_{x}\sin a\frac{k_{x}+\sqrt{3}k_{y}}{2}\sin a\frac{%
-k_{x}+\sqrt{3}k_{y}}{2}\right) ,
\end{equation}%
which is constructed based on the third-nearest-neighbor sites.

\subsubsection{$i$-wave magnets}

We use the tight-binding model for the $i$-wave magnets,%
\begin{equation}
\frac{2^{10}}{3\sqrt{3}}J\sigma _{z}\prod\limits_{j=0}^{2}\sin \left( \frac{a%
}{2}\mathbf{n}_{j}^{\text{A}}\cdot \mathbf{k}\right)
\prod\limits_{j=0}^{2}\sin \left( \frac{\sqrt{3}a}{2}\mathbf{n}_{j}^{\text{B}%
}\cdot \mathbf{k}\right)
\end{equation}%
It is constructed based on the 4th-nearest-neighbor sites. It is the
simplest tight-binding model with the $i$-wave symmetry because the
4th-nearest-neighbor sites are the minimum, which constitutes of the hopping
of 12 identical sites. Note that it is similar to the tight-binding model
used in the previous literature\cite{GI,Planar,MTJ,APEX,NLSeebeck} but
different, where the previous one is%
\begin{equation}
16J\sigma _{z}\prod\limits_{j=0}^{2}\sin \left( a\mathbf{n}_{j}^{\text{A}%
}\cdot \mathbf{k}\right) \prod\limits_{j=0}^{2}\sin \left( \sqrt{3}a\mathbf{n%
}_{j}^{\text{B}}\cdot \mathbf{k}\right) ,
\end{equation}%
which is constructed based on the 12th-nearest-neighbor sites.

\subsection{Topological numbers}

\subsubsection{Chern number}

\label{ChernN}

We show that the in-plane chiral ($p_{x}+ip_{y}$)-wave superconductor
coupled with altermagnets is a topological superconductor. The BdG
Hamiltonian reads%
\begin{equation}
H_{\text{BdG}}\left( \mathbf{k}\right) =%
\begin{pmatrix}
\hat{\xi}(\mathbf{k}) & \hat{\Delta}(\mathbf{k}) \\ 
\hat{\Delta}^{\dagger }(\mathbf{k}) & -\hat{\xi}^{T}(-\mathbf{k})%
\end{pmatrix}%
.
\end{equation}%
For the in-plane chiral $p$-wave superconductor in the presence of the
altermagnetic order, it is explicitly given by 
\begin{equation}
H_{\text{BdG}}\left( \mathbf{k}\right) =%
\begin{pmatrix}
\frac{\hbar ^{2}k^{2}}{2m}+Jf_{X}(\mathbf{k})-\mu & \Delta _{p}\left(
k_{x}+ik_{y}\right) \\ 
\Delta _{p}\left( k_{x}-ik_{y}\right) & -\frac{\hbar ^{2}k^{2}}{2m}-Jf_{X}(%
\mathbf{k})+\mu%
\end{pmatrix}%
\end{equation}%
in the continuum limit. We expand it in terms of the Pauli matrices as%
\begin{equation}
H_{\text{BdG}}\left( \mathbf{k}\right) =\mathbf{h}\left( \mathbf{k}\right)
\cdot \mathbf{\sigma },
\end{equation}%
where%
\begin{align}
h_{x}\left( \mathbf{k}\right) =& \Delta _{p}k_{x},\quad h_{y}\left( \mathbf{k%
}\right) =\Delta _{p}k_{y}, \\
h_{z}\left( \mathbf{k}\right) =& \frac{\hbar ^{2}k^{2}}{2m}+Jf_{X}(\mathbf{k}%
)-\mu .
\end{align}%
The Chern number is calculated as\cite{Hsi}%
\begin{equation}
C=\frac{1}{4\pi }\int \mathbf{h}\left( \mathbf{k}\right) \cdot \left( \frac{%
\partial \mathbf{h}\left( \mathbf{k}\right) }{\partial k_{x}}\times \frac{%
\partial \mathbf{h}\left( \mathbf{k}\right) }{\partial k_{y}}\right) d^{2}k,
\label{Chern}
\end{equation}%
which gives $C=1$. In the main text, we numerically evaluate the Chern
number without taking the continuum limit.

\subsubsection{Winding number}

\label{Wind}

We show that the out-of-plane $p_{x}$-wave superconductor coupled with
odd-parity magnets is a topological superconductor. The BdG Hamiltonian
reads 
\begin{equation}
H_{\text{BdG}}\left( \mathbf{k}\right) =%
\begin{pmatrix}
\hat{\xi}(\mathbf{k}) & \hat{\Delta}(\mathbf{k}) \\ 
\hat{\Delta}(\mathbf{k}) & -\hat{\xi}(\mathbf{k})%
\end{pmatrix}%
=\hat{\xi}(\mathbf{k})\tau _{z}+\hat{\Delta}(\mathbf{k})\tau _{x}.
\end{equation}%
The BdG Hamiltonian has chiral symmetry defined by%
\begin{equation}
\left\{ \tau _{y},H_{\text{BdG}}\left( \mathbf{k}\right) \right\} =0.
\end{equation}%
The chiral index is defined by%
\begin{equation}
W\equiv \frac{1}{2\pi }\int_{0}^{2\pi }\tau _{y}H_{\text{BdG}}^{-1}\left( 
\mathbf{k}\right) \frac{\partial _{k_{x}}H_{\text{BdG}}\left( \mathbf{k}%
\right) }{\partial k_{x}}dk_{x}.
\end{equation}%
By changing the basis, the BdG Hamiltonian is rewritten in the form of%
\begin{align}
H_{\text{BdG}}\left( \mathbf{k}\right) =& \hat{\xi}(\mathbf{k})\tau _{x}+%
\hat{\Delta}(\mathbf{k})\tau _{y}  \notag \\
=& 
\begin{pmatrix}
0 & \hat{\xi}(\mathbf{k})-i\hat{\Delta}(\mathbf{k}) \\ 
\hat{\xi}(\mathbf{k})+i\hat{\Delta}(\mathbf{k}) & 0%
\end{pmatrix}
\notag \\
=& 
\begin{pmatrix}
0 & q^{\ast }(\mathbf{k}) \\ 
q(\mathbf{k}) & 0%
\end{pmatrix}%
.
\end{align}%
Then, the chial index is identical to the winding number\cite{Ryu}%
\begin{equation}
W\left( k_{y}\right) \equiv \frac{1}{2\pi }\int_{0}^{2\pi }\frac{1}{q(%
\mathbf{k})}\frac{\partial q(\mathbf{k})}{\partial k_{x}}dk_{x}
\end{equation}%
with%
\begin{equation}
q(\mathbf{k})\equiv \hat{\xi}(\mathbf{k})+i\hat{\Delta}(\mathbf{k}).
\end{equation}%
The winding number is nonzero if $q(\mathbf{k})$ encircles the origin.

\subsubsection{$p$-wave magnets}

Eq.(\ref{qp}) is rewritten as%
\begin{equation}
q(\mathbf{k})=c(\mathbf{k})+M\mathbf{u}(\mathbf{k})
\end{equation}%
with the center 
\begin{equation*}
c(\mathbf{k})\equiv 2t\cos ak_{y}-\mu ,
\end{equation*}%
and%
\begin{align}
M\equiv & \left( 
\begin{array}{cc}
2t & J \\ 
0 & \Delta _{p}%
\end{array}%
\right) , \\
\mathbf{u}(\mathbf{k})=& \left( 
\begin{array}{c}
\cos ak_{x} \\ 
\sin ak_{x}%
\end{array}%
\right) .
\end{align}%
The orbit of $q(\mathbf{k})$ is given by the ellipse satisfying $c(\mathbf{k}%
)+M\mathbf{u}$ with $\left\vert \mathbf{u}\right\vert =1$. Then, the
condition that the ellipse encircles the origin is 
\begin{equation}
c(\mathbf{k})+M\mathbf{u}(\mathbf{k})=\mathbf{0}
\end{equation}%
for $\left\vert \mathbf{u}\right\vert <1$. We solve $\mathbf{u}$ as%
\begin{align}
\mathbf{u}(\mathbf{k})=& -M^{-1}c(\mathbf{k})  \notag \\
=& \left\{ \frac{J\Delta _{s}}{2t\Delta _{p}}+\mu -\cos ak_{y},-\frac{\Delta
_{s}}{\Delta _{p}}\right\} .
\end{align}%
Hence, the condition is obtained as%
\begin{equation}
\left( \frac{J\Delta _{s}}{2t\Delta _{p}}+\mu -\cos ak_{y}\right)
^{2}+\left( \frac{\Delta _{s}}{\Delta _{p}}\right) ^{2}<1.
\end{equation}%
The winding number is 1 when the ellipse encircles the origin.

\section*{Acknowledgement}

This work is supported by Grants-in-Aid for Scientific Research from MEXT
KAKENHI (Grant No. 23H00171).

\end{document}